\documentclass{amsart}
\usepackage{amsmath,amsthm,amsfonts,amssymb}
 \usepackage{graphicx}
 \usepackage{paralist}
  \usepackage{graphics} %% add this and next lines if pictures should be in esp format
  \usepackage{epsfig} %For pictures: screened artwork should be set up with an 85 or 100 line screen
\theoremstyle{definition}

\newtheorem{remark}{Remark}

\newcommand{\be}{ \begin{eqnarray} }
\newcommand{\ee}{ \end{eqnarray} }
\newcommand{\ben}{ \begin{eqnarray*} }
\newcommand{\een}{ \end{eqnarray*} }
\newcommand{\R}{ \mathbb{R} }

\newcommand{\bigM}{{\mathcal M}}
\newcommand{\bigF}{{\mathcal F}}
\newcommand{\real}{{\mathbb R}}

\title[Fokker-Planck Asymptotics for Traffic Flow Models]
      {Fokker-Planck Asymptotics for Traffic Flow Models}

\author[Michael Herty and Lorenzo Pareschi]{}


 \email{herty@mathc.rwth-aachen.de}
 \email{lorenzo.pareschi@unife.it}

\begin{document}
\maketitle

% Enter the first author's name and address:
\centerline{\scshape Michael Herty }
\medskip
{\footnotesize
% please put the address of the first author
 \centerline{RWTH Aachen University}
   \centerline{ D-52056 Aachen, Germany}
} % Do not forget to end the {\footnotesize by the sign }

\medskip

\centerline{\scshape Lorenzo Pareschi}
\medskip
{\footnotesize
 % please put the address of the second  and third author
 \centerline{ Department of Mathematics \& CMCS, University of Ferrara}
   \centerline{I-44100 Ferrara, Italy}
}

\bigskip

% The name of the associate editor will be entered by an editorial staff
% \centerline{(Communicated by the associate editor name)}

\date{November 19, 2009}

%The abstract of your paper
\begin{abstract}
Starting from microscopic interaction rules we derive kinetic
models of Fokker--Planck type for vehicular traffic flow. The
derivation is based on taking a suitable asymptotic limit of the
corresponding Boltzmann model. As particular cases, the derived
models comprise existing models.
 New Fokker--Planck models
are also given and their differences to existing models are
highlighted. Finally,  we report on numerical experiments.
\end{abstract}

\section{Introduction}
Kinetic models for traffic flow provide a description of vehicular
traffic based on spatial, temporal and velocity information and
have been subject to intense discussion in recent literature, see
e.g., \cite{He, KW97,KW981,KW982,KW00} and
\cite{IllnerKlarMaterne2003, IllnerStoicaKlarWegener}. In this
paper we contribute to this discussion in the following way: we
introduce microscopic interaction rules and derive existing and
new equations using formal asymptotics and limits. This approach
provides insight in the dominant terms of kinetic models and
shows possible extensions.
\par
Microscopic interaction rules yielding Boltzmann type traffic flow
equations have been analyzed in  \cite{HertyKlarPareschi2005}.
Therein, the models by Klar et. al. \cite{KW97,KW981,KW982,KW00}
could be derived from basic rules. However, the important class of
Fokker--Planck type traffic flow models as introduced by Illner
et. al. \cite{IllnerKlarMaterne2003} could not be obtained. The
present work now closes the remaining gap. We introduce general
microscopic interaction rules as well as a derivation leading to
the Fokker--Planck type models of Illner et. al. Further models
corresponding to modified interaction rules are also discussed. In
this work we do not want to discuss further the validity of such
Fokker-Planck models (see \cite{IllnerKlarMaterne2003} for a
detailed discussion). We only remark that these models are
particularly interesting from the computational viewpoint since
they keep the kinetic information on vehicles without requiring
the evaluation of the expensive velocity integrals as in
Boltzmann-like models.
\par
The main target is the single--lane,
  spatial homogenous Fokker--Planck type equations for traffic flow.
  The prototype example introduced ad--hoc in \cite{IllnerKlarMaterne2003} is given by
\begin{equation}\label{general} \partial_t f + \partial_v \left( B[f] f - D[f] \partial_v f
\right) = 0, \end{equation} where $f(v,t)$ is the density of
vehicles with velocity $v$ at time $t$. Its motivation  is as
follows: $B$ is a heuristically defined braking and acceleration
operator depending on moments of $f,$ i.e., macroscopic
quantities, like number density $\rho$ or mean velocity $u$. They
are defined as
\begin{eqnarray*}
\rho = \int_0^1 f(v) \, dv,\quad \rho u = \int_0^1 v f(v) \, dv.
\end{eqnarray*}
 The underlying  assumption on both operators is that drivers only
observe averaged quantities. The diffusive term is introduced to
model the fact that drivers may not observe the averaged moments
correctly. Furthermore, this operators allows for trivial
equilibria corresponding to synchronized flow. Examples for the
operators $B[f]$ and $D[f]$ are given in
\cite{IllnerKlarMaterne2003} and
\cite{HertyIllnerKlarPanferov2006}
\begin{eqnarray}\label{Illnerfp}
\nonumber
&&\partial_t f + \partial_v \Bigl( ( -c^I_B q(\rho,u,v) \rho (v-u)^2 \chi_{\{v>u\}} + c^I_A ( 1 - \rho) (v-u)^2 \chi_{\{ v<u\} } ) f \Bigr)
=\\[-.2cm]
\\[-.2cm]
\nonumber &&\partial_v \{ ( \sigma(\rho,u) |v-u|^{\gamma} )
\partial_v f \}.
\end{eqnarray}
Herein $c^I_B,c^I_A$ are positive constants corresponding to
braking and acceleration, $\chi$ is the indicator function.
 For convenience we normalize density and velocity such that
$\rho_{max} = u_{max}=1.$  The function $\sigma(\rho,u)$ ensures
that the boundary conditions in the stationary case are met. It
vanishes for large and small values of
 $\rho$ and $u$.  An explicit form can be found e.g. in
\cite{IllnerKlarMaterne2003}. It can be changed to
$\sigma(\rho,u)|v-u|^\gamma +\epsilon$, where $\epsilon$ is a
residual diffusion to prevent discontinuous equilibria, see
\cite{HertyIllnerKlarPanferov2006}. Furthermore, the function
$q(\rho,u)$ is motivated by additional lane--changing activity. It
is modeled as
$$q(\rho,u,v) = 1 - \left( \frac{  v -u }{ 1 - u } \right) ^{\delta}.$$
The coefficients $\gamma$ and $\delta$ satisfy $0< \gamma+\delta < 3$ in order to obtain well-defined steady states.
\par
In this paper, starting from microscopic considerations, we obtain
a microscopic motivation for the given form of
braking/acceleration and diffusion operators in the Fokker-Planck
model. The microscopic rules will also lead to different but still
qualitatively similar Fokker--Planck type models. Hence, in the
sequel we start from a microscopic model with precise interaction
rules between corresponding cars and derive on this grounds
kinetic models of Boltzmann type. A suitable asymptotic limit then
permits to recover the desired Fokker--Planck type models. We give
the presentation for a general interaction of rule and state the
corresponding equations for several well-known examples. Finally
we report on numerical results for the different
Fokker-Planck models.

\section{Microscopic interaction rules and Boltzmann models}

The interaction rules in traffic flow are different from the usual
gas-dynamics case. In particular, traffic ''particles'' are
anisotropic, since a car that drives behind another one and speed
up does not necessarily make the person in front go faster or
slower. Additionally,  acceleration and braking scenarios are not
necessarily symmetric to each other.
\par
We will present the derivation of the kinetic models for a general
microscopic interaction rule. The leading car is assumed to have
a pre--interaction velocity $v$ and a post--interaction velocity
$v'$. We distinguish two different scenarios: if the leading car
is slower than the one in front of it the leading car
accelerates, cf. (\ref{accel}). On the other hand, if it is faster
it will decelerate or brake, cf. (\ref{decel}). A simple model is
as follows: we compare our velocity $v$ with the velocity $w$ of the
car in front. If we are faster, we brake according to the
relative velocity, if we are slower we accelerate. Since we cannot
precisely estimate the leading car's velocity and we will made
mistakes when braking and accelerating. The modeling includes therefore
some
additional parameters and a noise term proportional to braking and
acceleration intensity characterized by a random variable $\xi$.

%One purposes is to obtain the Fokker--Planck model
%(\ref{Illnerfp}) from these microscopic interaction rules. In
%order to achieve this it turns out that we need to put a general
%power $\kappa $ in the noise.

The general form of the microscopic interaction rule is
\begin{subequations}\label{eq:mir}
 \begin{eqnarray}
\label{accel}  \mbox{Acceleration: } \qquad   v' = v + {A} ( V_A - v ) + \xi\nu(v) (V_A-v)^{\kappa}, \;  v < W \\
\label{decel}  \mbox{Braking: }    \qquad         v' = v -{B}
(v-V_B) +  \xi \nu(v)(v- V_B)^{\kappa}, \;  v > W
\end{eqnarray}
\end{subequations}
where $W \in (0,1)$ represents the estimate on the leading car velocity $w$
made by the driver. The coefficients
  $V_A \in (v,1]$, $V_B\in [0,v)$
represent the desired velocities when drivers accelerate and
brake.  Thus the natural choice is $W=W(w)$ with $W(\cdot)$ a
monotone non decreasing function of $w$. Beside the obvious choice
$W=w$ other choices for $W$, although less realistic, are
interesting since they lead to simplified set of equations. For
example $W=u$ where drivers are able to estimate only the mean
velocity of the flux. Similarly $V_A=V_A(v,W)$ and $V_B=V_B(v,W)$
may depend on the actual velocity $v$ and the estimated relative
velocity $W$.
\par
To complete the interaction we leave the velocity of $w$ of the car in front  unchanged, i.e.,
$$ w' = w.$$

In (\ref{eq:mir}) $A$ and $B$ are non negative constants in
$(0,1)$, $\kappa \geq 1$ is a power that calibrates the dependence
of the noise from the braking/acceleration dynamic, $0\leq
\nu(v)\leq 1$ is a function that takes into account that noise is
also proportional to the actual driver speed and $\xi$ is a random
variable with zero mean and variance $\sigma^2$ distributed
accordingly to $\eta(\xi)$.

The above microscopic relations include as particular cases
several well-known models. We give some examples.

\begin{itemize}

\item {\bf Example 1:} The model by Illner, Klar \& al. \cite{IKLUW99}.
In this model we assume that individual estimates correctly the
leading car's velocity and that acceleration and braking are
bounded by the maximal and minimal velocity instead of the
velocity of the leading car\\
\[
W=w,\,V_A=1,\,V_B=0,\,\nu(v)=1,\,k=1.
\]

\item {\bf Example 2:} A generalized model by G\"unther, Klar \&
al.\cite{KMWG01}. Drivers interact via symmetric braking and
acceleration scenarios governed by the leading car and can
estimate correctly the leading car velocity
\[
W=w,\,V_A=w,\,V_B=w.
\]
Note that in the original model no diffusion was present
$\nu(v)=0$.

\item {\bf Example 3:} If we assume that individuals react with respect to the
average velocity and that acceleration and braking are bounded by
the average velocity we have the set of parameters\\
\[
W=u,\,V_A=u,\,V_B=u.
\]
This model as we will see permits to derive the Fokker-Planck
equation (\ref{Illnerfp}) from the microscopic dynamics in a
suitable asymptotic scaling.
\end{itemize}

Some remarks are in order.
\begin{remark}~\rm\\
\begin{enumerate}
\item In the microscopic model it is possible to add a further term independent of the other drivers behavior
 \begin{eqnarray}
\label{diff}   v' = v  + {E} \left( \psi(v) - v\right).
\end{eqnarray}

From a modeling point of view this corresponds to free flow
traffic adjusting towards a certain desired velocity distribution
$\psi(\cdot).$  To keep the derivation simple we will
ignore the presence of this free flow term in the sequel.

\item
Note that in (\ref{eq:mir}) we require that the post-interaction
velocities satisfy $v'\in (0,1)$. We can show that this is
guaranteed by our assumptions on the parameters. Let us show this
for the acceleration term. Since $0\leq V_A-v \leq 1$ we have
\begin{eqnarray*}
v'&\leq& v+A(V_A-v)+|\xi|\nu(v)(V_A-v)\\
&=& v(1-A-|\xi|\nu(v))+(A+|\xi|\nu(v))V_A\\
 &\leq& V_A
\end{eqnarray*}
if $|\xi|\leq 1-A$. Note that for $A$ close to $1$ the domain of
$\xi$ vanishes. One can show that independently of $A$ we have
$v'\leq 1$ if $|\xi|\leq (1-V_A)/V_A$. However in the sequel we
will be interested in the behavior of the model for small values
of $A$ and $B$ and for such reason we will take simply $|\xi|\leq
1-A$.

It is easy to see that $v'\geq v$ if $|\xi|\leq A$. Note however
that in this case for small values of $A$ the domain of $\xi$
vanishes. If we assume that $\nu(v)\leq v$ we get
\begin{eqnarray*}
v'&\geq& v+A(V_A-v)-|\xi|\nu(v)(V_A-v)\\
&\geq& v+A(V_A-v)-|\xi|v\\
&= & v(1-A-|\xi|)+A V_A\\
 &\geq& A V_A
\end{eqnarray*}
if $|\xi|\leq 1-A$ which permits to pass to the limit $A\to 0$
keeping $\xi$ finite.

Similarly for the braking term we have $v'\in (V_B,1)$ provided
that $|\xi|\leq 1-B$ and $\nu(v)\leq 1-v$. Thus if we want to
preserve the lateral bounds of $v$. To this end the presence
of the  function $\nu(v)$ is of importance. $\nu$ is such that it vanishes at
the extreme values of the velocity $v=0$ and $v=1$. This can be achieved by
any function such that $\nu(v)\leq \min\{v,1-v\}$ for $v\in
(0,1)$. For example we might take
\[
\nu(v)=\frac12(1-|2v-1|)\,
\]
or a quadratic function like \[\nu(v)=v(1-v).\] This implies that
the diffusion term in (\ref{accel})-(\ref{decel}) vanishes in
correspondence of the desired speeds $V_{A,B}$ and at the extremal
values $0$ and $1$ of the velocity domain.
\end{enumerate}
\end{remark}

As mentioned in the introduction a general Boltzmann--type
formulation for vehicular traffic was derived in
\cite{HertyKlarPareschi2005}. For simplicity we restrict
to the case $W=w$ in (\ref{eq:mir}) we will come back to other
choices at the end of the next section.

In weak form we can write for the car distribution $\tilde f(v,t)$
the kinetic equation
\begin{eqnarray}
\partial_t \int_0^\infty \tilde f(v) \phi(v) \, dv &=& Q_A + Q_B \label{fp1}
\end{eqnarray}
where
\begin{subequations}\label{weakform}
\begin{eqnarray}
Q_A &:=& \int_0^1 \int_v^1 \int_{\R} \beta^A_{ ({v},{w}) \to (v',w')}\tilde f(v) \tilde f(w) ( \phi(v') - \phi(v)) \, d\xi \, dw \, dv  \label{accelweakform}\\
Q_B &:=& \int_0^1 \int_0^v \int_{\R} \beta^B_{ ({v},{w}) \to
(v',w')} \tilde f(v) \tilde f(w) ( \phi(v') - \phi(v)) \, d\xi \,
dw \, dv. \label{decelweakform}
\end{eqnarray}
\end{subequations}
 Additionally, as in \cite{HertyKlarPareschi2005} we assume that the function $\beta$ is given by
 \begin{eqnarray*}
\beta^{A,B}_{ ({v},{w}) \to (v',w')} &=& \left\{
\begin{array}{rcrcl}
    \eta(\xi)|{v}-{w}| q_A\chi_{[0,1]}(v') && {v}& < & {w} \\
    \eta(\xi)|{v}-{w}| q_B\chi_{[0,1]}(v') && {v}& > & {w},
 \end{array} \right.
\end{eqnarray*}
where $\chi$ is the indicator function. In the general case
(\ref{eq:mir}) the collision kernel has a weight of $|v-W|$.

From (\ref{accelweakform})-(\ref{decelweakform}) conservation of
the total number of vehicles is obtained for $\phi(v)=1$ which
represents the only conservation property satisfied by the
equation. If we consider the case $\phi(v)=v$ (this corresponds to
the behavior of the mean velocity) we obtain
\begin{eqnarray*}
\partial_t \int_0^\infty \tilde f(v) v \, dv =
A \int_0^1 \int_v^1 \int_{\R} \beta^A_{ ({v},{w}) \to
(v',w')}\tilde f(v) \tilde f(w) (V_A-v) \, d\xi \, dw \, dv+ \\
\int_0^1 \int_v^1 \int_{\R} \beta^A_{ ({v},{w}) \to (v',w')}\tilde
f(v) \tilde f(w) \xi\nu(v)(V_A-v)^k \, d\xi \, dw \, dv-\\
B \int_0^1 \int_0^v \int_{\R} \beta^B_{ ({v},{w}) \to
(v',w')}\tilde f(v) \tilde f(w) (v-V_B) \, d\xi \, dw \, dv+ \\
\int_0^1 \int_0^v \int_{\R} \beta^B_{ ({v},{w}) \to (v',w')}\tilde
f(v) \tilde f(w) \xi\nu(v)(v-V_B)^k \, d\xi \, dw \, dv.
\end{eqnarray*}
Hereby we assume that the random variables in the acceleration and
the braking terms take values on the sets $(A-1, 1-A)$ and $(B-1,
1-B)$ respectively. Now the second and the fourth integrals vanish
since $\xi$ has zero mean and so we have
\begin{eqnarray*}
\partial_t \int_0^\infty \tilde f(v) v \, dv =
A \int_0^1 \int_v^1 |v-w| \tilde f(v) \tilde f(w) (V_A-v) \,  dw \, dv-\\
B \int_0^1 \int_0^v |v-w| \tilde f(v) \tilde f(w) (v-V_B) \,  dw
\, dv,
\end{eqnarray*}
which represents the change in mean velocity due to acceleration
and braking.

If we assume $A=B=\gamma$ and $V_A=V_B=V$ we can write
\begin{eqnarray*}
\partial_t \int_0^\infty \tilde f(v) v \, dv =
\gamma \int_0^1 \int_0^1 |v-w| \tilde f(v) \tilde f(w) (V-v) \, dw
\, dv.
\end{eqnarray*}
In Example 2 we have $V_A=V_B=w$. In this case a straightforward
computation shows that the mean velocity $u$ is conserved. This
situation has much in common with a granular gas dynamic where
$\gamma$ is related to the coefficient of restitution, and the
large time behavior of the system in absence of diffusion
($\nu(v)=0$) is described by synchronized traffic given by a Dirac
delta $f(v)=\delta(v-u)$ where all cars have the same speed $u$.
Note that the synchronized traffic state is an equilibrium state
of the system even in the general case when $A\neq B$ and
$\nu(v)\neq 0$ since the noise term vanishes when all cars have
the same speed.
%More precisely one expects that for large values of $\sigma^2$
%(when compared to $A$ and $B$) the steady state is unstable
%whereas for smaller values of $\sigma^2$ cars converge towards
%synchronized traffic. Here we do not analyze further this aspect.

One of the major drawbacks of Boltzmann-type models is their
excessive complexity which makes it difficult the development of
numerical simulation methods as well of analytical tools for the
computation of steady states.

In order to obtain simpler kinetic Fokker--Planck type equations
we use a suitable asymptotic limit similar to the technique
presented in \cite{PareschiToscani2006}. In the latter
presentation only a {\em global} and symmetric interaction rules
for a Maxwellian-like kernel have been considered. Other similar
interaction rules in the context of opinion formation has been
studied in \cite{Toscani}. We emphasize that in the following, the
interaction is anisotropic and the collision kernel is of
hard-spheres type. Therefore it is different to
\cite{PareschiToscani2006, Toscani} and also to standard
approaches in gas--dynamics.
\par

\section{Fokker--Planck type asymptotic models}

First let us recall some definitions. We define $\bigM_0$ the
space of all probability measures in $\real_+$ and by
\be\label{misure} \bigM_{p} =\left\{ \eta \in\bigM_0:
\int_{\real_+} |\xi|^{p}\eta(\xi)\, d\xi < +\infty, p\ge
0\right\}, \ee the space of all Borel probability measures of
finite momentum of order $p$, equipped with the topology of the
weak convergence of the measures. Let $\bigF_p(\real_+)$, $p>1$ be
the class of all real  functions on $\real_+$ such that $g(0)=
g'(0) =0$, and $g^{(m)}(v)$ is H\"older continuous of order
$\delta$,
 \[\label{lip} \|g^{(m)}\|_\delta= \sup_{v\not= w} \frac{|g^{(m)}(v) -g^{(m)}(w)|}{
|v-w|^\delta} <\infty,
 \]
 the integer $m$ and the number $0 <\delta \le 1$ are such that $m+\delta =p$, and
$g^{(m)}$ denotes the $m$-th derivative of $g$.

 In the rest of the paper we will assume that the symmetric probability density $\eta(\xi)$
 which characterizes the transition rate  belongs to $\bigM_{2+\alpha}$, for some $\alpha
 >0$.  Moreover, to simplify  computations, we assume that this density is obtained
 from a given random variable $Y$ with zero mean and unit
 variance, that belongs to $\bigM_{2+\alpha}$. Thus, $\eta$ of variance $\sigma^2$ is the density of
 ${\sigma}Y$. By this assumption, we can easily obtain the dependence on $\sigma$ of the moments of $\eta$.
 In fact, for any $p >2$ such that the $p$-th moment of $Y$ exists,
 \[
\int_{\real}|\xi|^{p}\eta(\xi) d\xi = E\left(
\left|{\sigma}Y\right|^{p}\right) =
\sigma^{p}E\left(\left|Y\right|^{p}\right).
\]

To skip inessential difficulties, that do not change the following
analysis, we suppose that the random variables in the acceleration
and the braking terms take values on the sets $(A-1, 1-A)$ and
$(B-1, 1-B)$ respectively. We also assume that
$\phi\in\bigF_{2+\delta}(\real_+)$ for $\alpha
 \geq \delta>0$.

In order to derive a Fokker--Planck equation from (\ref{weakform})
we need to consider an appropriate scaling of the equations. We
set for $\epsilon>0$
\begin{equation}\label{scaling}
\tau = \epsilon t, \; f(v,\tau ) = \tilde f( v, t).
\end{equation} and {assume } that there exists non--negative constants $c_{A}, c_{B}$ and $c_{D}$ with the following properties
\begin{eqnarray}
\lim_{A \to 0,\,\epsilon \to 0}  \frac{A}\epsilon = c_{A} , \qquad
\lim_{B \to 0,\,\epsilon \to 0}  \frac{B}\epsilon = c_{B},\qquad
\lim_{\sigma \to 0,\,\epsilon \to 0} \frac{\sigma^2}\epsilon =
c_{D}. \label{scaling constants}
\end{eqnarray}
These assumptions corresponds to an asymptotic situation where
each interaction produces very small velocity variations and in
the limit we preserve the main features (deterministic
acceleration/braking and random noise) of the dynamic.

The scaled kinetic equation reads
\begin{eqnarray}\label{scaled kinetic}
\partial_\tau \int_0^\infty f(v) \phi(v) \, dv &=& \frac{1}{\epsilon} ( Q_A(f,\phi,\eta) + Q_B(f,\phi,\eta) ). \label{fpeq2}
\end{eqnarray}

For small values of $A, B$ and $\sigma^2$ we have $v' \approx v$
and so we can consider a Taylor expansion in the collision kernels
$Q_{A}$ and $Q_{B}$ of $\phi(v')$ around $v.$ Let us consider here
the acceleration term (\ref{accel})
\begin{eqnarray*}
 \phi(v') - \phi(v) =(A (V_A-v) + \xi\nu(v) (V_A-v)^\kappa) \phi_v(v) + \\
  \frac{1}{2} ( A (V_A-v) + \xi\nu(v) (V_A-v)^\kappa )^2 \phi_{vv}(\tilde
  v),
\end{eqnarray*}
where for some $\theta \in (0,1)$ we have
\[
\tilde v = (1-\theta)v + \theta v'.
\]

Now we insert this expansion into the acceleration operator $Q_A$
and compute the limit for $\epsilon \to 0.$ Since $\eta(\xi)$ has
mean zero and vanishing variance we get
\begin{eqnarray*}
 \lim_{\epsilon \to 0 } \frac{1}{\epsilon}  Q_A( f,\phi,\eta)
&=& q_{A} \left\{c_A\int_0^1 \int_v^1  (w-v)(V_A-v) f(v) f(w) \phi_v(v)  \, dw \, dv.\right.\\
&+&\left.\frac{1}{2} c_{D} \int_0^1 \int_v^1
(w-v)\nu(v)^2(V_A-v)^{2\kappa} f(v) f(w)  \phi_{vv}(v) \, dw \,
dv\right\}\\
&+&\lim_{\epsilon \to 0 } \frac{1}{\epsilon}R(f,\phi,\eta)
\end{eqnarray*}
where
\begin{eqnarray*}
R(f,\phi,\eta)&=&\frac{1}{2} q_A c_{D} \int_0^1
\int_v^1\int_{\real}
\eta(\xi)(w-v)\nu(v)(A (V_A-v) + \xi\nu(v) (V_A-v)^\kappa)^{2}\cdot\\
&&\cdot f(v) f(w) (\phi_{vv}(\tilde v)-\phi_{vv}(v))\,d\xi\, dw \,
dv.
\end{eqnarray*}
Since $\phi \in \bigF_{2+\delta}(\real_+)$, and $|\tilde v - v |=
\theta|v'-v|$
 \be\label{rem}
 \left| \phi_{vv}(\tilde v)- \phi_{vv}( v)\right| \le \| \phi_{vv}\|_\delta |\tilde v - v |^\delta \le
 \| \phi_{vv}\|_\delta |v' - v |^\delta .
  \ee
  Hence
 \begin{eqnarray*}
|R(f,\phi,\eta)| &\le& q_A c_D\frac{\|
\phi_{vv}\|_\delta}{2}\int_0^1 \int_v^1\int_{\real}
\eta(\xi)(w-v)\nu(v)\cdot
\\&& \cdot | A (V_A-v) + \xi\nu(v) (V_A-v)^\kappa |^{2+\delta} f(w)f(v)\,d\xi\,d
w\,d v.
 \end{eqnarray*}
By virtue of the inequality
 \[
 |A (V_A-v) + \xi\nu(v) (V_A-v)^\kappa|^{2+\delta} \le C_\delta
( A^{2+\delta}+|\xi|^{2+\delta}),
 \]
 and using the fact that $v$ is compactly supported in $(0,1)$
we finally obtain the bound
 \be\label{to0}
|R(f,\phi,\eta)| &\le& q_A c_D\frac{\| \phi_{vv}\|_\delta}{2}
C_\delta\rho^2
\left(A^{2+\delta}+\int_{\real}\eta(\xi)|\xi|^{2+\delta}\,d\xi\right)
 \ee
 Since $\eta$ is a probability density with zero mean and
$\sigma^2$ variance, and $\eta$ belongs to $\bigM_{2+\alpha}$,
for  $\alpha
 >\delta$,
 \[
\int_{\real}|\xi|^{2+\delta}\eta(\xi) d\xi = E\left(
\left|{\sigma}Y\right|^{2+\delta}\right) =
\sigma^{2+\delta}E\left(\left|Y\right|^{2+\delta}\right),
\]
and $E\left(\left|Y\right|^{2+\delta}\right)$ is bounded. Using
this equality one shows that $\frac{1}{\epsilon}R(f,\phi,\eta)$
converges to zero as $\epsilon \to 0$.

Thus we obtain
\begin{eqnarray*}
 \lim_{\epsilon \to 0 } \frac{1}{\epsilon}  Q_A( f,\phi,\eta)
&=& q_{A} \left\{c_A\int_0^1 \int_v^1  (w-v)(V_A-v) f(v) f(w) \phi_v(v)  \, dw \, dv.\right.\\
&+&\left.\frac{1}{2} c_{D} \int_0^1 \int_v^1
(w-v)\nu(v)^2(V_A-v)^{2\kappa} f(v) f(w)  \phi_{vv}(v) \, dw \,
dv\right\}
\end{eqnarray*}

By similar arguments for $Q_{B}$ we get
\begin{eqnarray*}
 \lim_{\epsilon \to 0 } \frac{1}{\epsilon}  Q_B( f,\phi,\eta)
&=& q_{B} \left\{c_B\int_0^1 \int_0^v  (v-w)(V_B-v) f(v) f(w) \phi_v(v)  \, dw \, dv.\right.\\
&+&\left.\frac{1}{2} c_{D} \int_0^1 \int_0^v
(v-w)\nu(v)^2(V_B-v)^{2\kappa} f(v) f(w)  \phi_{vv}(v) \, dw \,
dv\right\}.
\end{eqnarray*}
In order to present the strong formulation of the limiting kinetic
equation we introduce the following constants
\begin{equation} \tilde c_{A}  = c_{A} q_{A} , \; \lambda_{A} = c_{D} q_{A}, \; \lambda_{B} = c_{D} q_{B},\; \tilde  c_{B} = c_{B} q_{B}. \end{equation}
Summarizing, the above computations we observe that in the limit
$\epsilon \to 0$ we obtain the following Fokker--Planck equation
given in strong form by
\begin{eqnarray}
\label{FP full} \nonumber
\partial_{\tau} f + \partial_{v} \left\{ f \left(  \tilde c_{A}  \int_{v}^1 f(w) (w-v)(V_A-v) dw  - \tilde c_{B} \int_{0}^v f(w) (v-w)(v-V_B) dw
\right) \right\} = \\[-.25cm]
\\
\nonumber
 \frac{1}2 \partial_{vv} \left\{ \nu(v)^2 f \left(  \lambda_{A} \int_{v}^{1} f(w) (w-v)(V_A-v)^{2\kappa} dw + \lambda_{B}
 \int_{0}^v f(w) (v-w)(v-V_B)^{2\kappa}  dw \right) \right\},
\end{eqnarray}
which can be rewritten in more compact notation as
\begin{eqnarray}
\partial_{\tau} f + \partial_{v} \left( f(v)\int_{0}^1 |v-w| L(v,w)f(w)dw \right) = \\[-.25cm]
\\
\nonumber
 \frac{1}2 \partial_{vv} \left( \nu(v)^2 f(v)\int_{0}^{1} |v-w| D(v,w) f(w) dw
 \right),
\end{eqnarray}
where
\[
L(v,w)=\left\lbrace\begin{array}{cc}
                \tilde c_A(V_A-v) & v<w, \\
                \tilde c_B(V_B-v) & v>w, \\
              \end{array}
              \right.
              \qquad
D(v,w)=\left\lbrace\begin{array}{cc}
                \lambda_A|V_A-v|^{2\kappa} & v<w, \\
                \lambda_B|V_B-v|^{2\kappa} & v>w. \\
              \end{array}
              \right.
\]

Furthermore, it is useful to introduce the partial moments for
fast and slow cars as follows
\begin{eqnarray*}
 \rho_{S} = \int_{0}^v f(w)  dw, && \rho_{F} = \int_{v}^1 f(w)   dw, \\
 u_{S}= \frac{1}{\rho_S} \int_{0}^v f(w) w dw, && u_{F}= \frac{1}{\rho_F} \int_{v}^1 f(w) w   dw, \\
 T_{S} = \frac{1}{\rho_S} \int_{0}^v f(w) ( v- w)(v-V_B) dw, &&  T_{F} = \frac{1}{\rho_F} \int_{v}^1 f(w) ( w- v)(V_A-v) dw, \\
 W^k_{S} = \frac{1}{\rho_S} \int_{0}^v f(w)  (v-w)( v-V_B)^{2\kappa} dw, &&
 W^k_{F} = \frac{1}{\rho_F} \int_{v}^1 f(w) (w-v)( V_A- v)^{2\kappa}
 dw.
 \end{eqnarray*}
 \par
In this way we can rewrite the limiting Fokker-Planck equation in
the form
\begin{equation}\label{FP}
\partial_{\tau} f + \partial_{v} \left( f \left(  \tilde c_{A} \rho_F T_{F}  - \tilde c_{B} \rho_S T_{S} \right) \right) =
 \frac{1}2 \partial_{vv} \left( \nu(v)^2 f \left(  \lambda_{A} \rho_F  W^{\kappa}_{F} + \lambda_{B} \rho_S W^{\kappa}_{S} \right) \right)
\end{equation}

For the interaction rules presented in the previous section we
have the following equations in the limit $\epsilon \to 0.$

\begin{itemize}

\item {\bf Example 1:} For the model by Illner, Klar \& al. we get
\begin{eqnarray}
\nonumber
\partial_{\tau} f + \partial_{v} \left\{ f \left(  \tilde c_{A}  (1-v)\int_{v}^1 f(w) (w-v) dw  - \tilde c_{B} v \int_{0}^v f(w) (v-w) dw
\right) \right\} = \\[-.25cm]
\label{FP example 1}
\\
 \frac{1}2 \partial_{vv} \left\{ \nu(v)^2 f \left(  \lambda_{A} (1-v)^{2}\int_{v}^{1} f(w) (w-v) dw + \lambda_{B}
 v^{2}\int_{0}^v f(w) (v-w)  dw \right) \right\} \nonumber
\end{eqnarray}
or using the partial moments
\begin{eqnarray*}
\partial_{\tau} f + \partial_{v} \left( f \left(  \tilde c_{A}( 1- v) ( \rho_{F}u_{F} - \rho_{F} v)
  - \tilde c_{B}  v  ( \rho_{S} v - \rho_{S}u_{S} ) \right) \right) = \\
 \frac{1}2 \partial_{vv} \left( \nu(v)^2 f \left(  \lambda_{A} (1-v)^2  ( \rho_{F}u_{F} - \rho_{F} v)
 + \lambda_{B}  v^2 ( \rho_{S} v - \rho_{S}u_{S} ) \right) \right)
\end{eqnarray*}

\item {\bf Example 2:} For the generalized model by G\"unther,
Klar \& al. we obtain
\begin{eqnarray}\nonumber
\partial_{\tau} f + \partial_{v} \left\{ f \left(  \tilde c_{A}  \int_{v}^1 f(w) (w-v)^2 dw  - \tilde c_{B} \int_{0}^v f(w) (v-w)^2 dw
\right) \right\} = \\[-.25cm]
\label{FP example 2}
\\ \nonumber
 \frac{1}2 \partial_{vv} \left\{ \nu(v)^2 f \left(  \lambda_{A} \int_{v}^{1} f(w) (w-v)^{2+k} dw + \lambda_{B}
 \int_{0}^v f(w) (v-w)^{2+k}  dw \right) \right\}.
\end{eqnarray}

\item {\bf Example 3:} For the model where individuals react with respect to the
average velocity we have

\begin{subequations}\label{mh:03}
\begin{eqnarray}
\partial_{\tau} f + \partial_{v} \left( f \left(  \tilde c_{A}( v- u)^2 \rho \chi_{\{v<u\}}
  - \tilde c_{B}  (v-u)^2 \rho \chi_{\{v>u\}} \right) \right) = \\
 \frac{1}2 \partial_{vv} \left( \nu(v)^2 f \left(  \lambda_{A} (u-v)^{1+2\kappa}  \rho \chi_{\{v<u\}}
 + \lambda_{B}  (v-u)^{1+2\kappa} \rho \chi_{\{v>u\}} \right) \right)
\end{eqnarray}
\end{subequations}
\end{itemize}

%\item[III]  Using the same derivation as in case I we obtain
%\begin{eqnarray*}
%\partial_{\tau} f + \partial_{v} \left( f \left(  \tilde c_{A}( 1- v) ^\alpha( \rho_{F}u_{F} - \rho_{F} v)
%  - \tilde c_{B}  v^\alpha  ( \rho_{S} v - \rho_{S}u_{S} ) \right) \right) = \\
% \frac{1}2 \partial_{vv} \left( f \left(  \lambda_{A} (1-v)^{2\beta}  ( \rho_{F}u_{F} - \rho_{F} v)
% + \lambda_{B}  v^{2\beta} ( \rho_{S} v - \rho_{S}u_{S} ) \right) \right)
%\end{eqnarray*}
%\end{enumerate}

\begin{remark}~\rm
\begin{itemize}
\item The evaluation of the partial moments which appears
in examples 1 and 2 can be done efficiently at a numerical level
at the same computational cost $O(N)$ of the computation of mass
and momentum. This is of paramount importance in applications of
the the Fokker-Planck models to realistic simulations.
\item The general Fokker-Planck equation (\ref{FP full}) derived above preserve the
essential properties of the microscopic relations it came from.
For example mean velocity is preserved only when $V_A=V_B=w$ and
$A=B$. Similarly synchronized traffic is a possible equilibrium
state of the system when $V_A=V_B=w$. Moreover upper and lower
bounds for the cars velocities are satisfied thanks to the
presence of $\nu(v)$.
\end{itemize}
\end{remark}

\section{Relations with Illner, Klar, Materne model}

We analyze some of the derived models and compare them to Illner,
Klar, Materne model.

Let us at first consider Example 3 and assume that $\nu(v)$ is
independent of $v$ and set
$$\lambda_{A} \equiv \lambda_{B} =: \lambda/\nu^2,$$
i.e.,  $q_{A}=q_{B}$.

We can differentiate in the diffusion term and obtain
\begin{eqnarray*}
\partial_{\tau} f +
\partial_{v} \Bigl( f ( ( \tilde c_{A} (v-u)^2  + \frac{(1+2\kappa)\lambda}2  ( u-v )^{2\kappa} ) \rho
\chi_{\{v<u\}}
\\
  - ( \tilde c_{B}  (v-u)^2 +  \frac{ (1+2\kappa) \lambda}2   (v-u)^{2\kappa} ) \rho \chi_{\{v>u\}} )  \Bigr) =
 \frac{1}2 \partial_{v} \left( \partial_{v} f |u-v|^{1+2\kappa} \lambda \rho
 \right).
\end{eqnarray*}
By comparing this equation with the original model by Illner et.
al. (\ref{Illnerfp}) we observe the following. Since we did not
include lane changing in the microscopic interaction rules we can
only compare with (\ref{Illnerfp}) in the case $P=0.$ The
 previous derivations hold true also in the case
 when the quantities $\tilde c_{A}, \tilde
c_{B}$ and $\lambda$ are $\rho$ and $u$ dependent. Hence, if
they are chosen  such that
\begin{eqnarray} \label{constants}
\tilde c_{A} + \frac{1+2\kappa}2 \lambda  =  c^I_{A} \frac{ 1-\rho
}\rho,  \quad \tilde c_{B} + \frac{1+2\kappa}2 \lambda  = c^I_{B},
\quad
 \lambda  =  \sigma(\rho,u) \rho,  \quad 1+ 2 \kappa = \gamma
 \end{eqnarray}
 we exactly recover the model by lllner et. al. (\ref{Illnerfp}) for $\gamma=3.$ In the cases $\gamma \not = 3$ we can
 recover the exponent in the diffusion but obtain additional drift terms to a power of $\gamma-1.$
  We also observe that we immediately obtain the lower bound on $\gamma>1$ for positive $\kappa.$

Moreover the diffusion term is degenerate when $v=u$. Adding a
modified diffusion such as $\sigma(\rho,u)+\epsilon$ will off
course remove the degeneracy but destroys the possibility of
synchronized traffic as discussed in
\cite{HertyIllnerKlarPanferov2006}.

Let us now consider Example 2.
  We approximate the partial moments in (\ref{FP example 2}) close to synchronized traffic
$$ f = \delta ( v- u).$$
At synchronized traffic the operators $T_{S}$ and $W_{S}$ collapse to
\begin{eqnarray*}
T_{S} = \frac{1}{\rho}  ( u-v)^2 \chi_{\{v>u\}}, && T_{F} = \frac{1}{\rho}  ( u-v)^2 \chi_{\{v<u\}}, \\
W_{S} = \frac{1}{\rho}  ( u-v)^{1+2\kappa} \chi_{\{v>u\}}, &&
W_{F} = \frac{1}{\rho}  ( u-v)^{1+2\kappa} \chi_{\{v<u\}},
\end{eqnarray*}
and under the additional assumption $\lambda_{A} \equiv
\lambda_{B} =: \lambda$ and using the same choices for the
constants $\tilde c_{A}, \tilde c_{B}$ and $\lambda$ as before, we
again obtain (\ref{Illnerfp}) in the case $P=0$. Therefore, the
model (\ref{FP example 2}) can be seen as a generalized equation
comprising (\ref{Illnerfp}) and at the same time linking the
coefficients appearing in diffusion and braking terms to
microscopic interaction rules.

Note however that in the computation reported above to connect our
Fokker-Planck models with Illner \& al. model we assumed that
$\nu(v)$ is constant. Thus there is no mechanism to prevent the
solution to violate the lateral bounds for the car velocity.

We discuss some particular cases. The limit case in Illner's model
for the diffusion coefficient is $\gamma=1$ which implies
$\kappa=0$ in the microscopic rules (\ref{eq:mir}). This implies a
constant diffusion in the microscopic interaction rules
(\ref{eq:mir}). This seems at least questionable. On the other
hand taking $\kappa=1$ corresponds to $\gamma=3$ being the upper
limit case for Illner' s model. In this case the noise is directly
proportional to the difference in the velocities.

Finally, we compare the diffusion terms of (\ref{Illnerfp}), i.e.,
$$ \partial_{v }  \left(  \partial_{v} f  \sigma(\rho,u)  |u-v|^\gamma  \right) $$
and (\ref{FP example 2}) in the case $\lambda_{A} = \lambda_{B} = \lambda$, i.e.,
$$ \partial_{v} \left(  \partial_{v} f \frac{\lambda(\rho,u)}2  \int_{0}^1 f(w) |w-v|^{1+2\kappa} dw \right).$$
We observe, that  the Fokker--Planck derived from microscopic
rules has a convolution integral in the diffusion term compared to
the degenerated diffusion in the Illner model. Hence, in the
steady states we expect a smoother distribution profile $f(v).$
Furthermore, the degeneracy at $v=u$ is not present in the second
case. This is a remarkable feature of (\ref{FP example 2}) with
respect to (\ref{Illnerfp}).

%%%%%%%%%%%%%%%%%%%%%%%%%%%%%%%%

\section{Steady states and numerical experiments}

In this section we report on numerical results for the models
(\ref{FP example 1}), (\ref{FP example 2}) and (\ref{mh:03}). In
all models we set the cut--off in the diffusive part as $\nu(v) = v
(1-v)$ and hereby normalize the computational domain for $v$ to $(0,1).$ We use
$N_{v}=100$ discretization points on $(0,1)$ for all computations.
We apply the methods of lines with a time horizon of $T=250$. The
large time horizon is chosen in order to capture the steady state
solution. When discretizing in $v$ we use a simple first--order
upwind discretization for the first--order derivatives and
centered differences for the diffusion term. The final system of
ordinary differential equations is then solved by an implicit
Euler method. During simulations the total mass $\rho = \int f dv$
is preserved.  The initial data for the simulation is chosen such
that it vanishes at the boundary and  given by $ f_{0} = \exp( -
25 \left( v - \frac{1}2 \right)^2 ).$ The initial mass is
$\rho_{0} \approx 0.35.$ For simplicity  we set $\lambda =
\lambda_{A}=\lambda_{B}$ in all simulations.
\par
\begin{figure}[htbp]
   \centering
   \epsfig{file=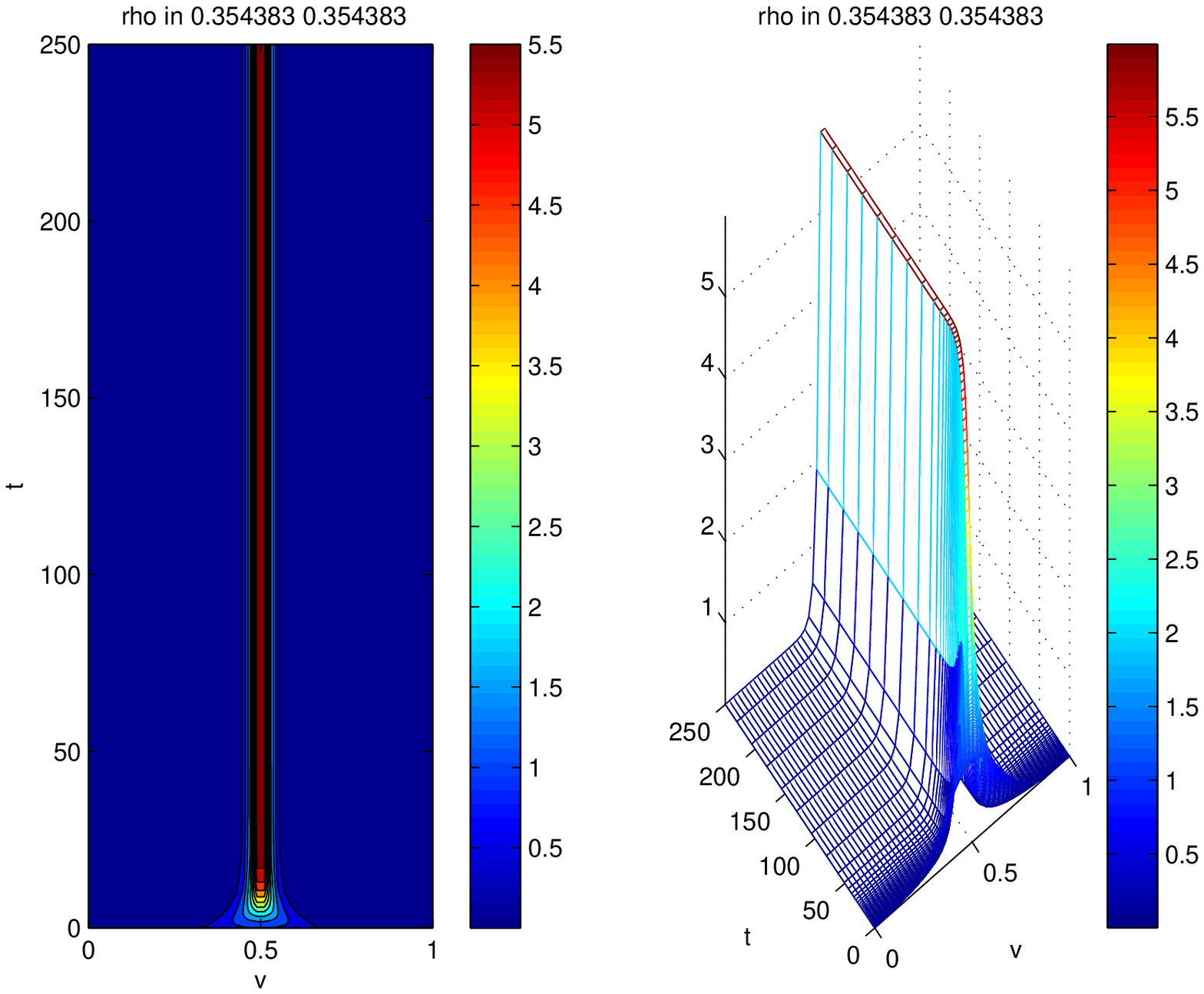, width=.45\textwidth} % requires the graphicx package
   \epsfig{file=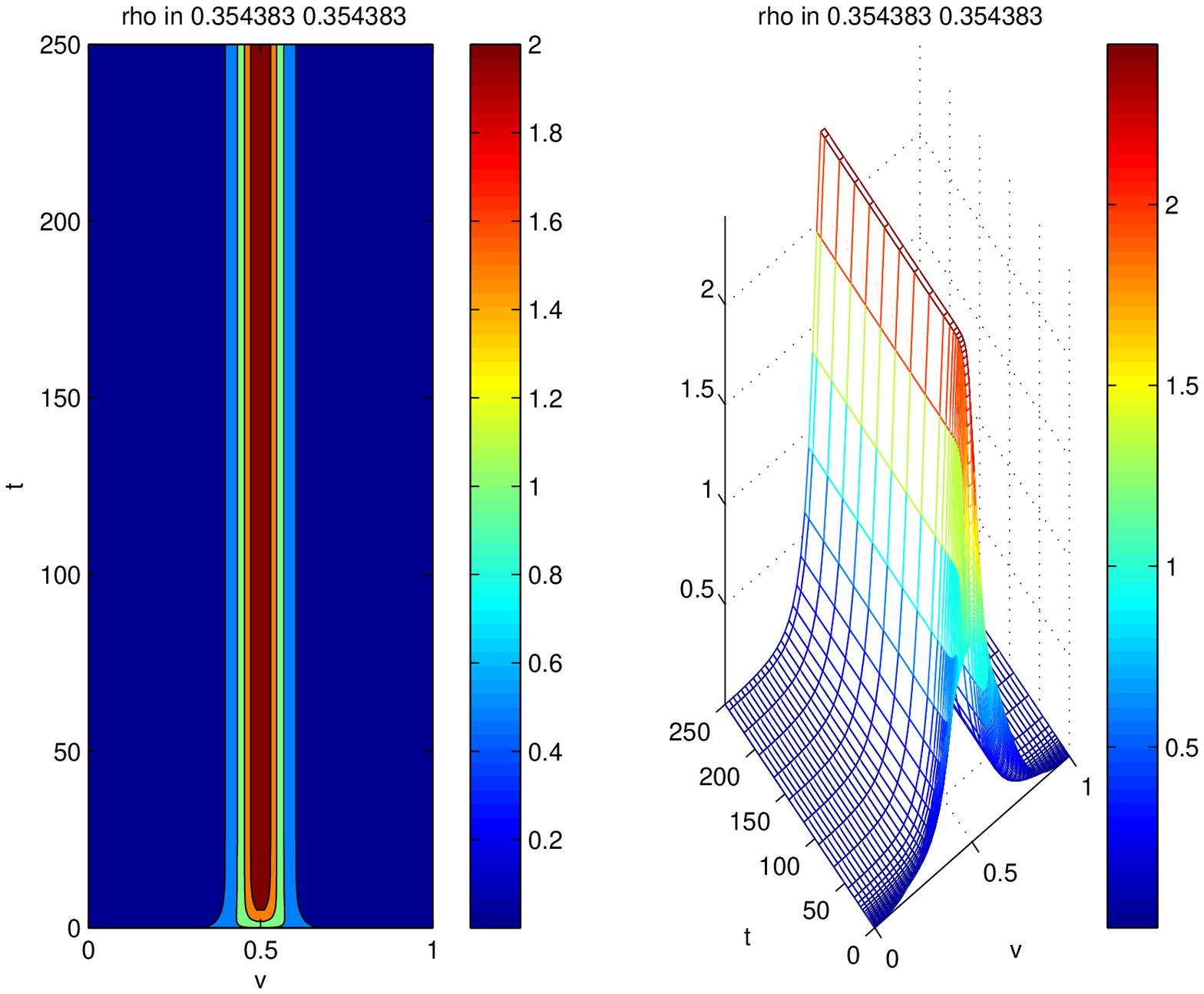,  width=.45\textwidth} \\% requires the graphicx package
   \epsfig{file=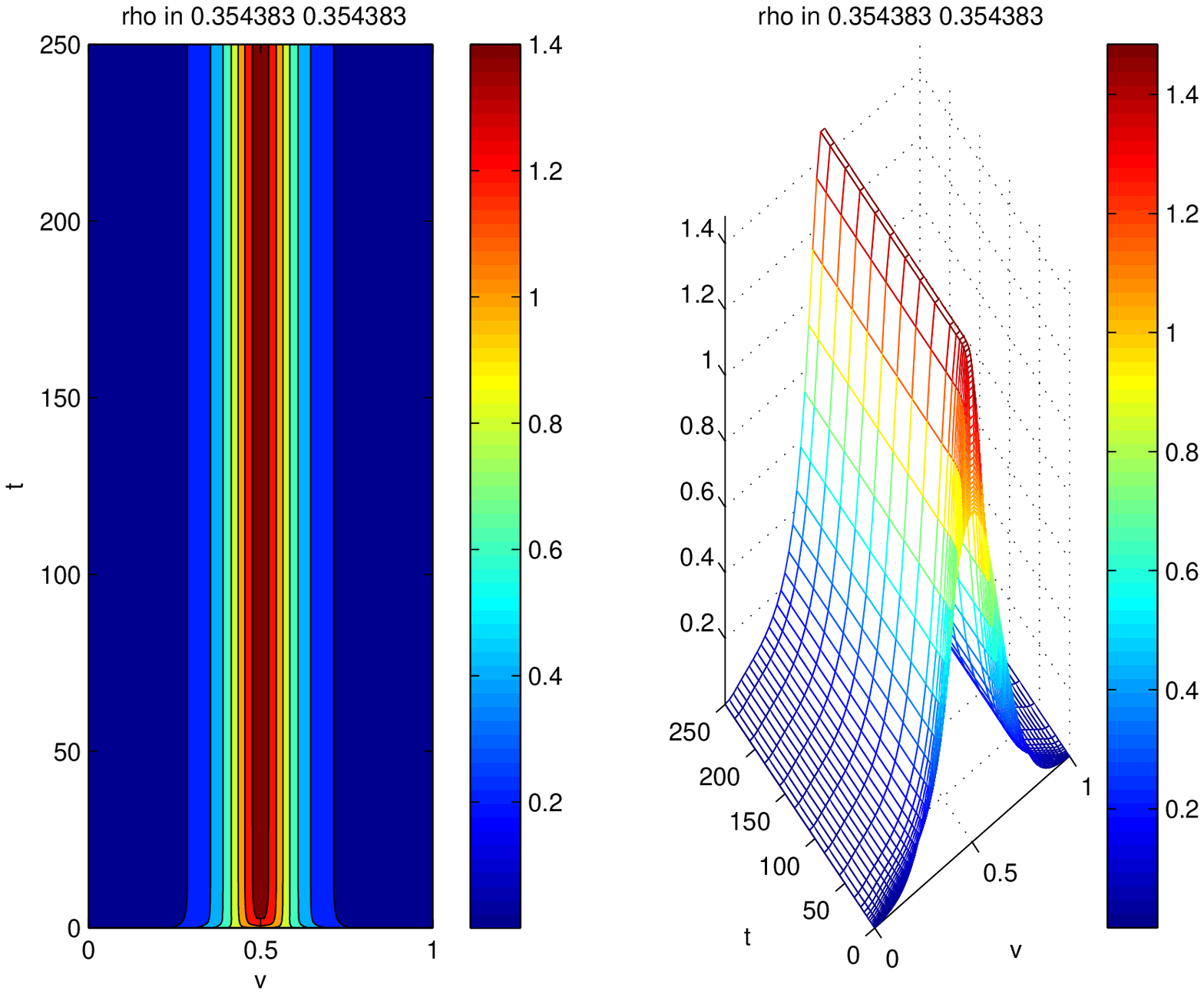,  width=.45\textwidth} % requires the graphicx package
   \epsfig{file=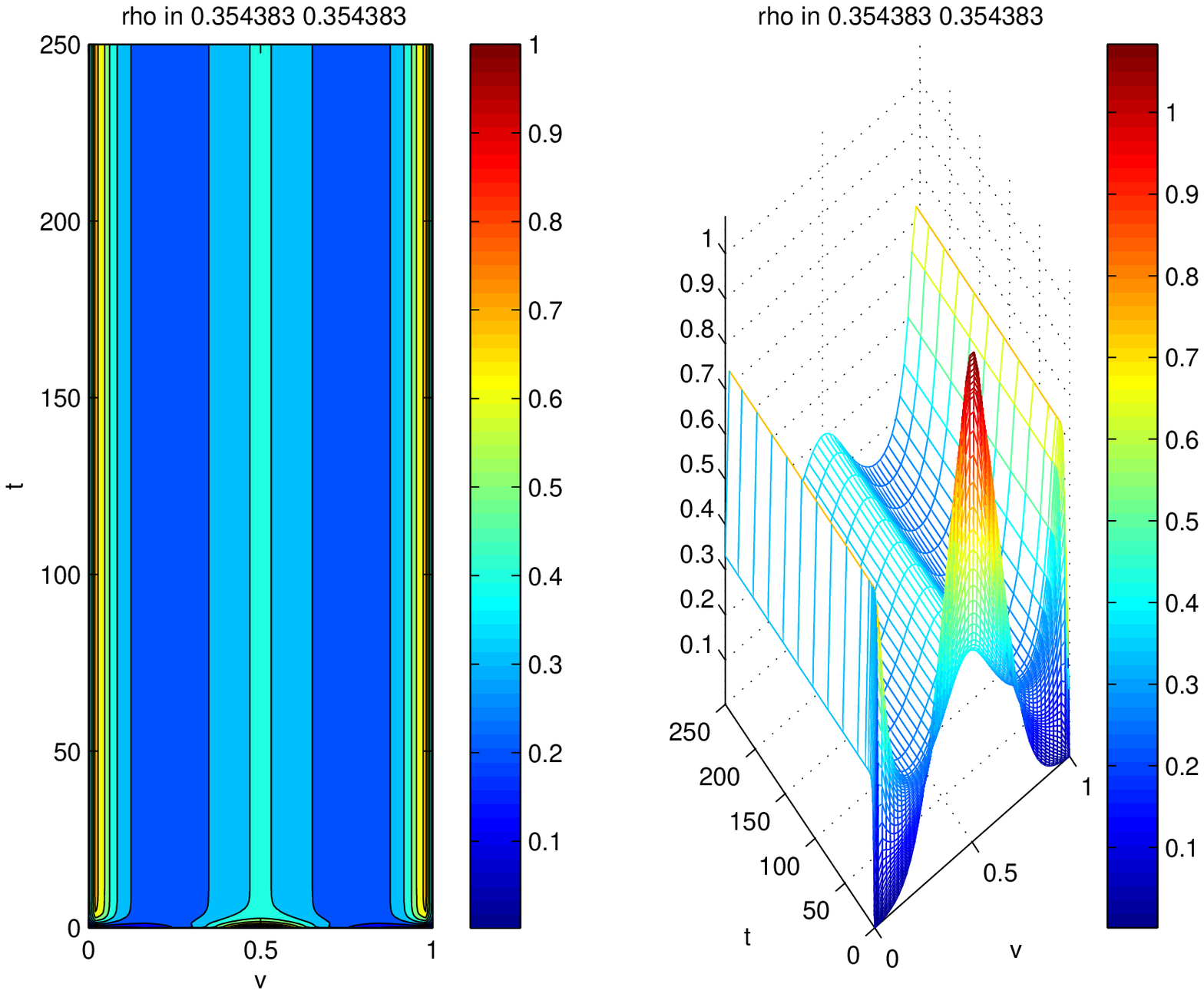,  width=.45\textwidth} % requires the graphicx package
  \caption{Example 1, numerical results for a diffusion coefficient of $\lambda=1$ (top left), $\lambda=5$ (top right), $\lambda=10$ (bottom left) and $\lambda=50$ (bottom right).}
\label{f:1}
\end{figure}
\begin{figure}[htbp]
   \centering
   \epsfig{file=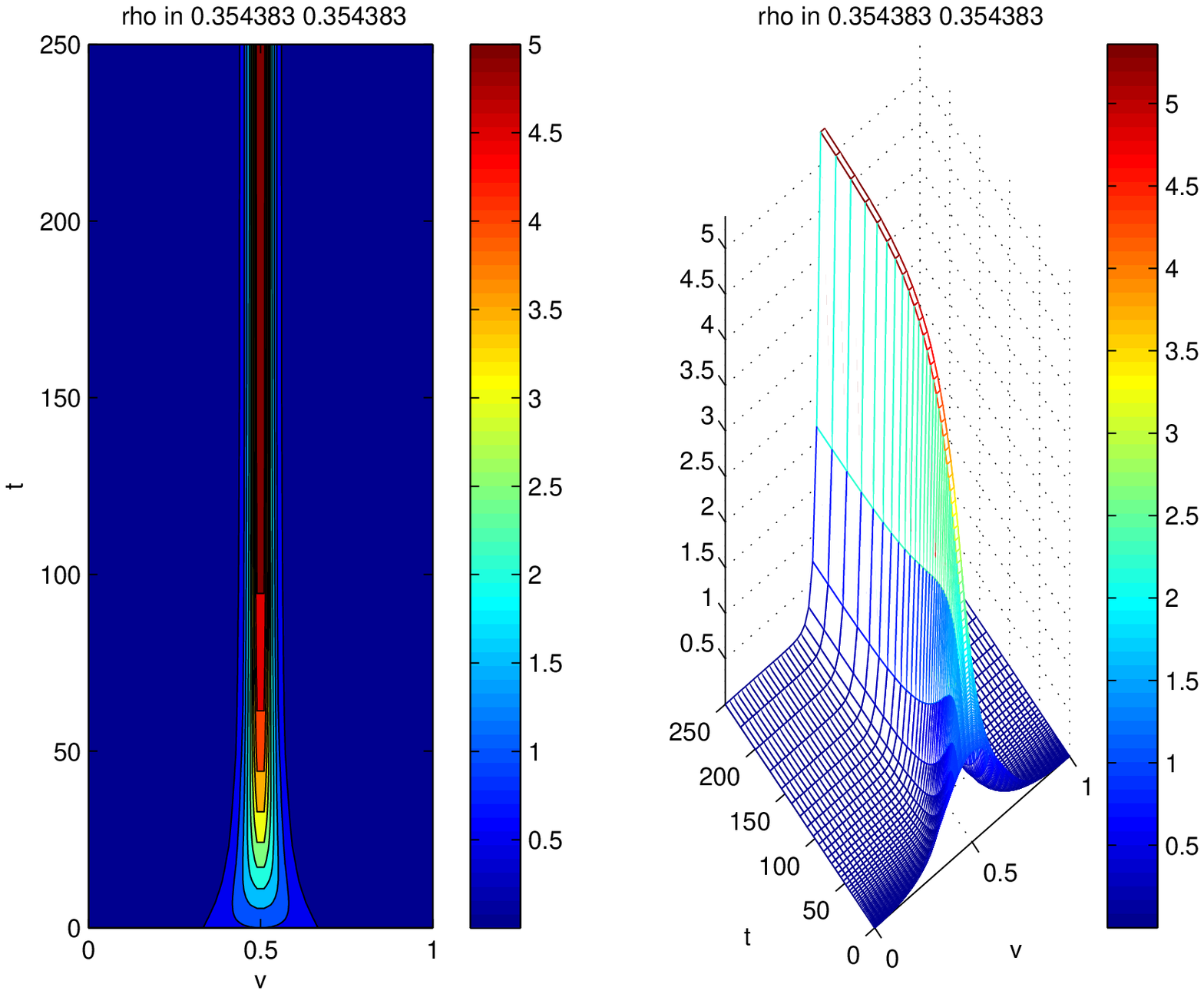, width=.45\textwidth} % requires the graphicx package
   \epsfig{file=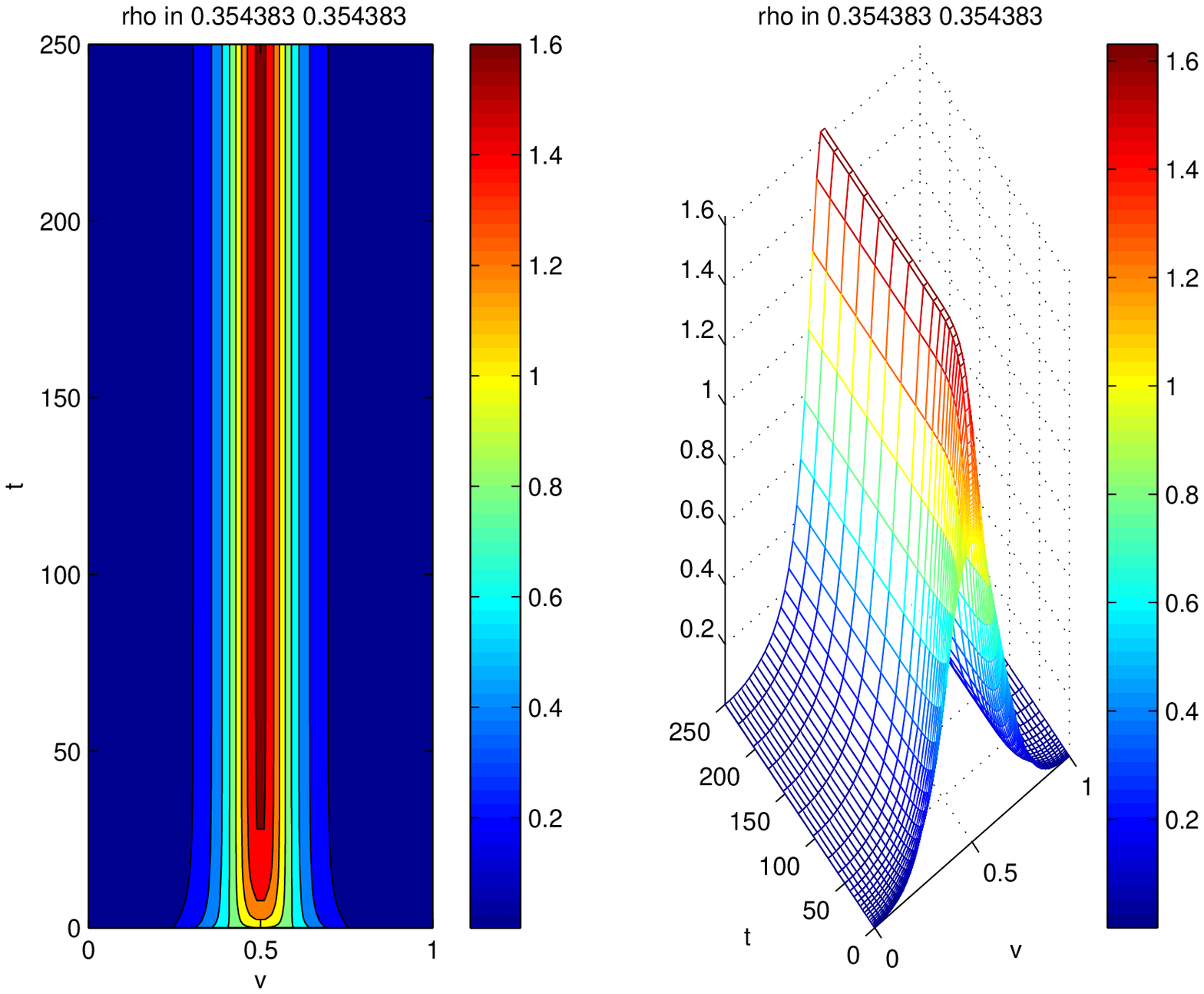,  width=.45\textwidth} \\% requires the graphicx package
   \epsfig{file=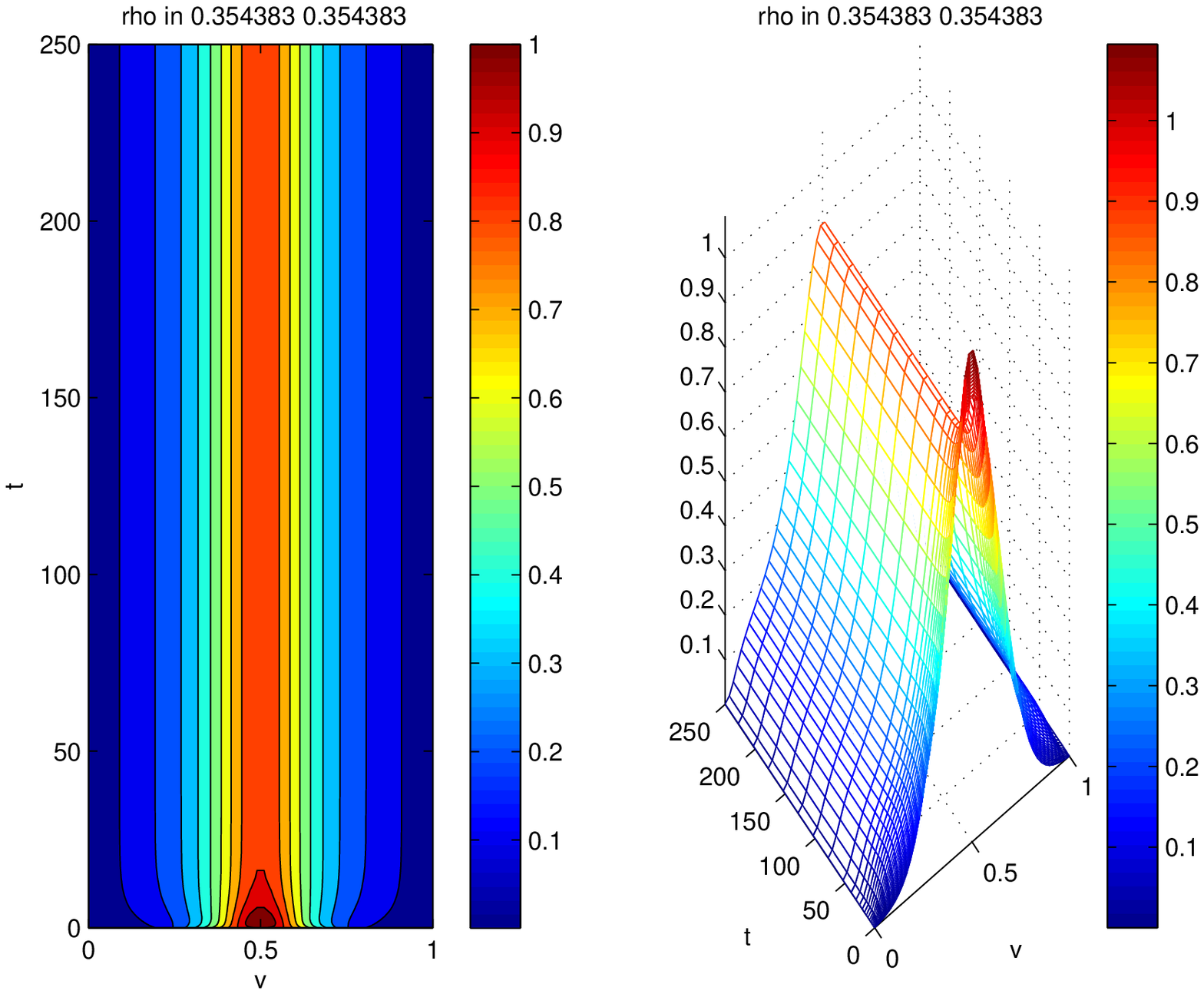,  width=.45\textwidth} % requires the graphicx package
   \epsfig{file=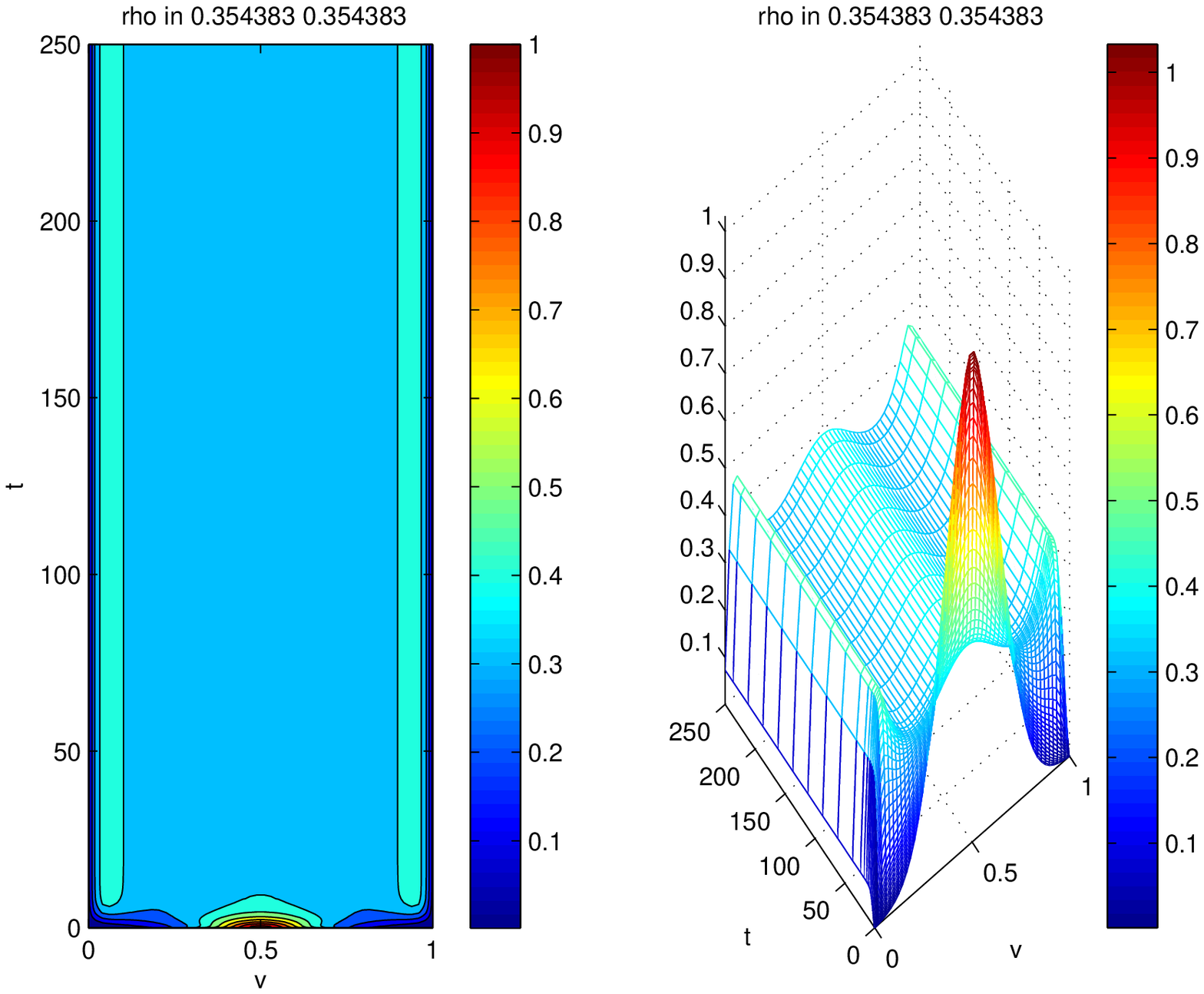,  width=.45\textwidth}\\ % requires the graphicx package
   \epsfig{file=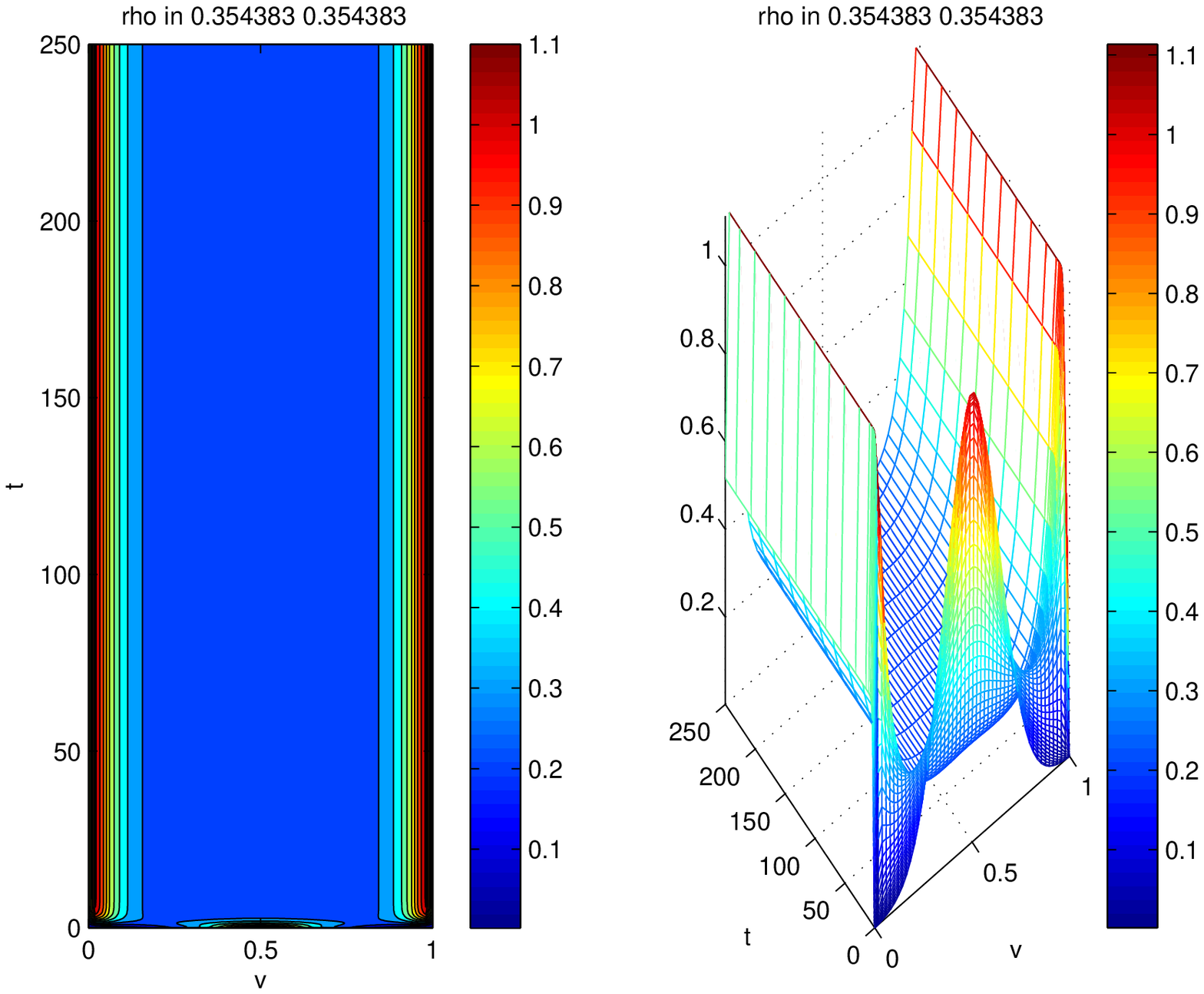,  width=.45\textwidth} % requires the graphicx package
    \caption{Example 2, numerical results for a diffusion coefficient of $\lambda=1$ (top left), $\lambda=5$ (top right), $\lambda=10$ (middle left) and $\lambda=25$ (middle right) and $\lambda=50$ (bottom). }
\label{f:2}
\end{figure}
\begin{figure}[htbp]
   \centering
   \epsfig{file=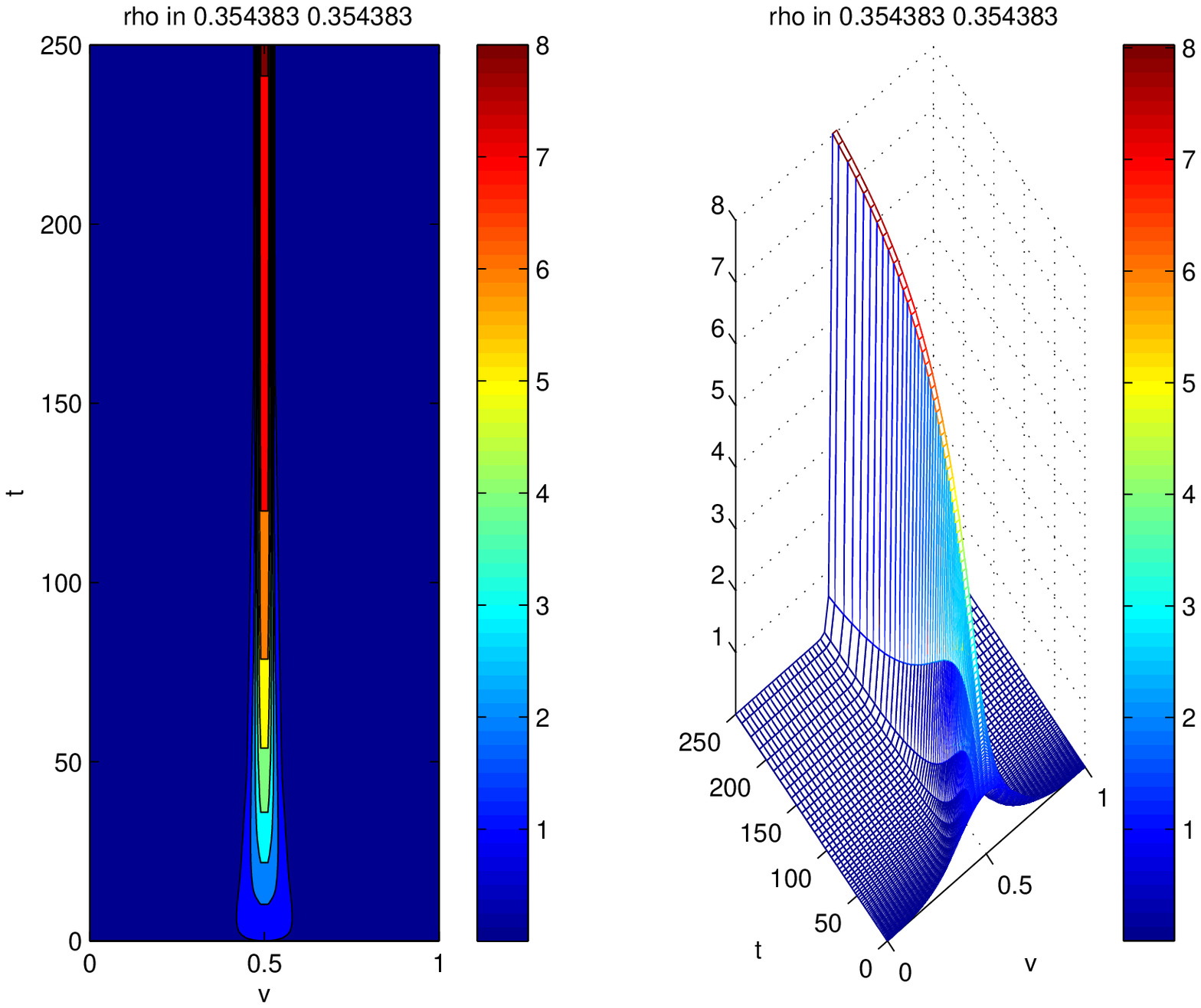, width=.45\textwidth} % requires the graphicx package
   \epsfig{file=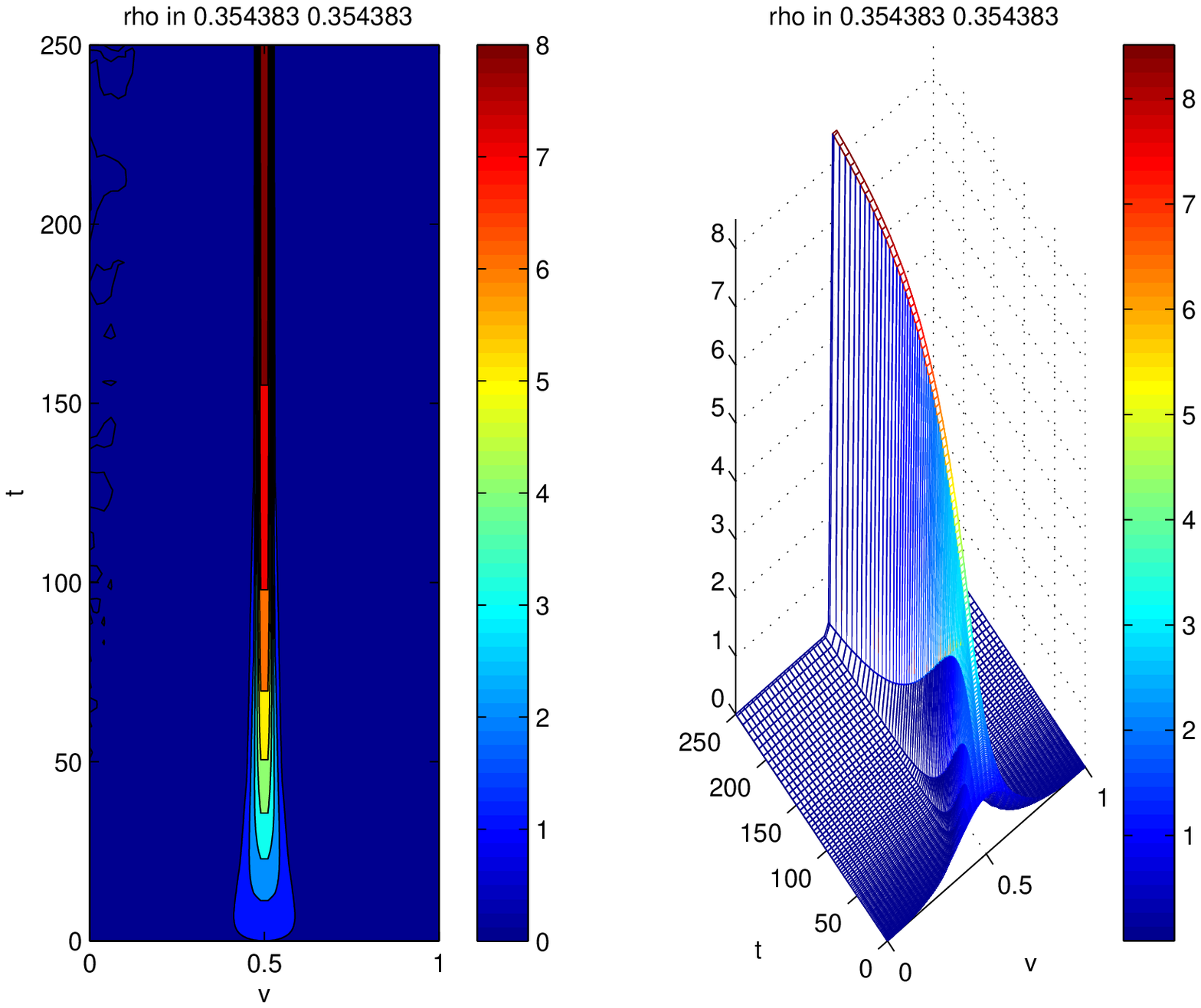,  width=.45\textwidth} \\% requires the graphicx package
    \caption{Example 2, numerical results for different interaction coefficients  $\kappa=1$ (left) and  $\kappa=2$ (right). }
\label{f:3}
\end{figure}
\begin{figure}[htbp]
   \centering
   \epsfig{file=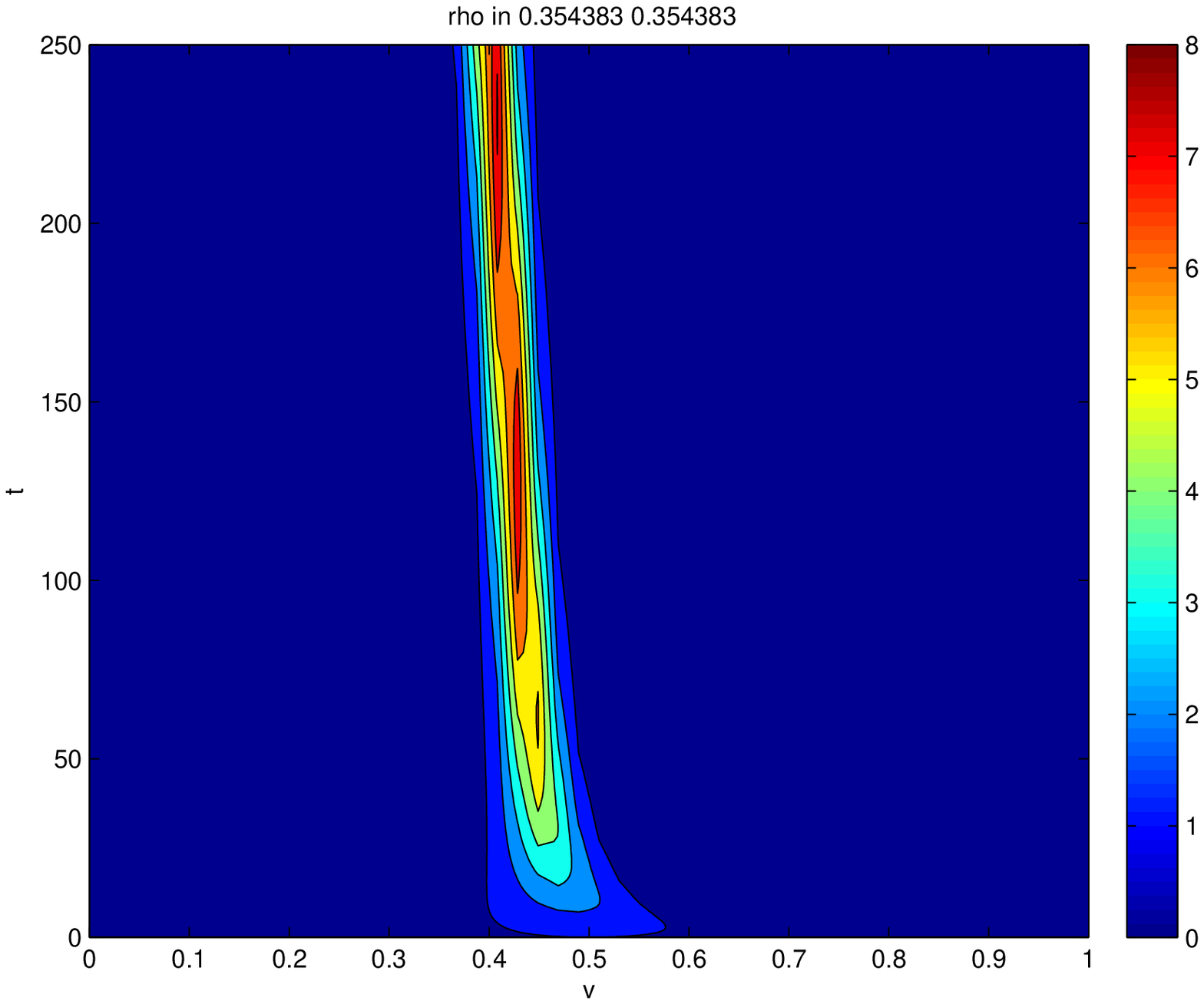, width=.45\textwidth} % requires the graphicx package
   \epsfig{file=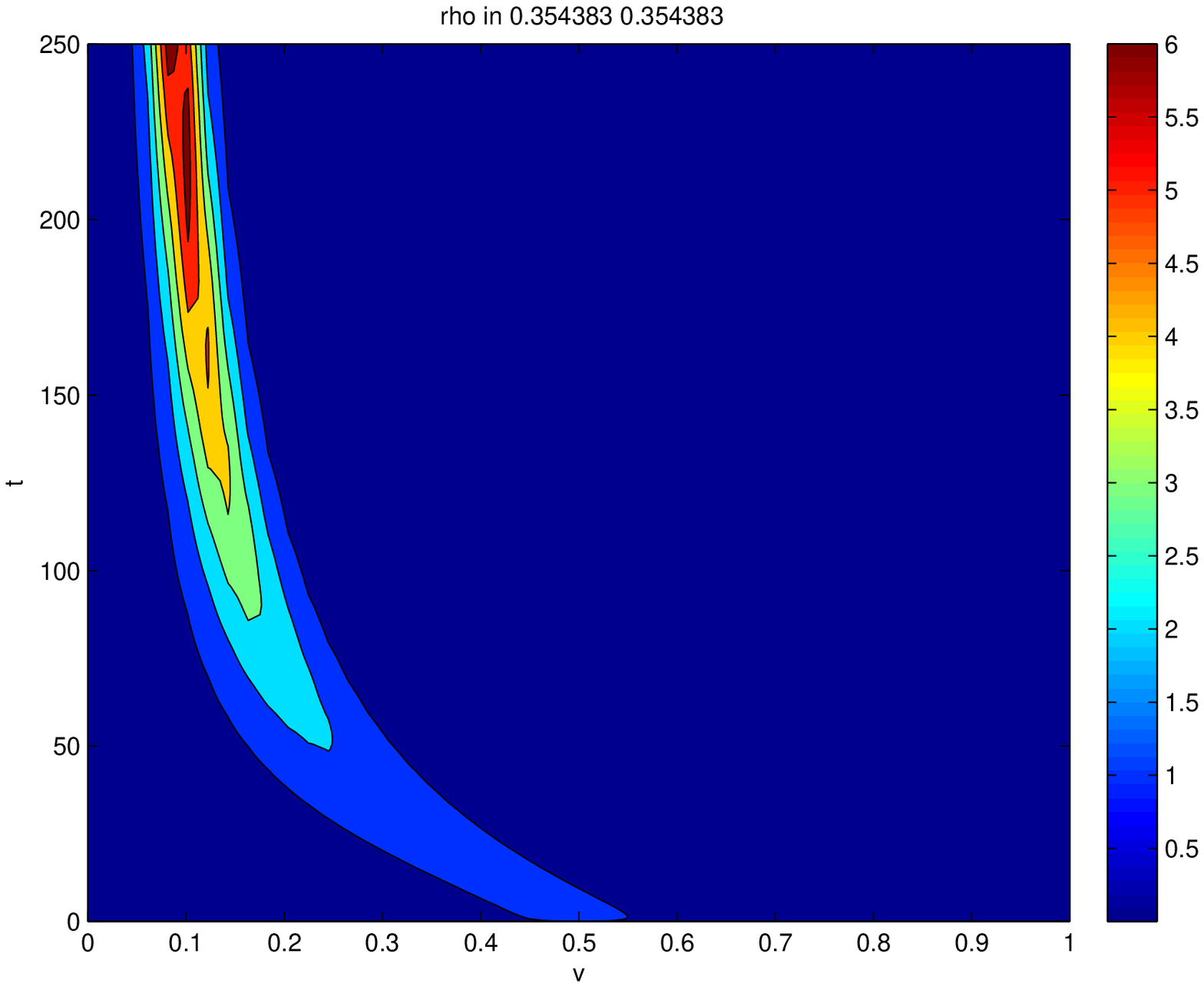,  width=.45\textwidth} % requires the graphicx package
    \caption{Example 2, numerical results for $c_{B} /   c_{A}=5$ (left) and $c_{B} /   c_{A}=10$ (right).  }
\label{f:4}
\end{figure}

At first we study the dependence of the time evolution of $f_{0}$ on the diffusion coefficient $\lambda$ if the dynamics is controlled by example 1 (\ref{FP example 1}), i.e.,
\begin{eqnarray*}
\partial_{t} f + \partial_{v} \left( f c_{A} (1-v) \int_{v}^1 f(w)(w-v) dw - c_{B} v \int_{0}^v f(w) (v-w) dw \right) = \\
\frac{1}2 \partial_{vv} \left\{ \nu(v)^2 f \lambda \left( (1-v)^2
\int_{v}^1 f(w) ( w-v) dw + v^2 \int_{0}^v f(w)(v-w) dw\right)
\right\}.
\end{eqnarray*}
The braking and acceleration constants are $c_{B}=c_{A}=1$. We
present plots of the isolines of the solution as well as a
three--dimensional for diffusion coefficients between $\lambda=1$
and $\lambda=50$ in figure \ref{f:1}. We observe  a
concentration forming from the initial Gaussian curve. As
we increase the diffusion parameter the center of the
concentration moves from $v=\frac{1}2$ to the boundaries $v=\{0, 1\}$.
 A similar behavior is observed when
simulating example 2, (\ref{FP example 2}), i.e.,
\begin{eqnarray*}
\partial_{t} f + \partial_{v} \left( f c_{A} \int_{v}^1 f(w)(w-v)^2 dw - c_{B} \int_{0}^v f(w) (v-w)^2 dw \right) = \\
\frac{1}2 \partial_{vv} \left(  \nu(v)^2 f \lambda \int_{0}^1 f(w)
| w-v|^{2+\kappa} dw  \right).
\end{eqnarray*}
Using the same parameters  and setting $\kappa=0$ we observe in figure \ref{f:2}  the same qualitative behavior. For small diffusion coefficients we observe a concentration at the center $v=\frac{1}2$ and for larger coefficients at the boundary $v=0$ and $v=1$. Next, we study the dependence of the time evolution on the other parameters present in example 2. In figure \ref{f:3} we observe that the exponent $\kappa$ in (\ref{FP example 2}) does not influence the qualitative behavior. We give a result for a moderate diffusion parameter and the case $\kappa=1$ and $\kappa=2.$ However, the braking and acceleration constants have a strong influence on the position of the concentration. In the case $\kappa=1$ and $\lambda=5$ and $\frac{c_{B}}{c_{A}} > 1$ we observe a shift of the point of concentration away from $v=\frac{1}2$.  Clearly, the obvious interpretation is: if the braking force is stronger than the acceleration force the equilibrium distribution must concentrate in a regime of lower velocity. The results are given in figure \ref{f:4} and has to be compared with the corresponding part in the upper left of figure \ref{f:1} where we used $\frac{c_{B}}{c_{A}}=1.$

Concerning Example 3, (\ref{mh:03}), a comparison between the derived model and the original model
of Illner et. al. is given.  Here, we compare the long term behavior in terms of the arising steady state.
In case of Illner's model (\ref{Illnerfp}) these can be computed explicitly \cite{HertyIllnerKlarPanferov2006}
and are given by
\begin{equation}\label{steady illner}
f(v) = \left\{
\begin{array}{ll}
    \displaystyle c^1 \exp \left(\frac{-c^I_{B}\rho}{ (3-\gamma)\sigma(\rho,u)} (v-u)^{3-\gamma}   \right), & v>u \\
    \displaystyle c^2 \exp \left(\frac{-c^I_{A}(1-\rho) }{ (3-\gamma)\sigma(\rho,u)} (u-v)^{3-\gamma}   \right), & v<u. \\
\end{array}\right.
\end{equation}
Here, it is assumed that $P=0$. The constants $c^{1,2}$ are such that mass and momentum
equation are satisfied, i.e., $\rho = \int f dv, \rho u = \int f v dv.$ In case of Example 3 we also
compute the steady states as
\begin{equation*}
f(v) = \left\{
\begin{array}{ll}
    \displaystyle
\frac{ c^1 }{ \nu^2(v) } \exp\left( \left( - \frac {\lambda_{B} \rho   }{ 2 + 2\kappa } ( v-u )^{2+2\kappa} \right) \right) \exp\left( - \rho \int_{u}^v \frac{ c_{B}}{ \nu^2(\eta) } ( \eta-u)^2 d\eta \right) & v>u,\\
\displaystyle \frac{ c^2 }{ \nu^2(v) } \exp\left( \left( -  \frac
{\lambda_{A} \rho   }{ 2 + 2\kappa } ( u -v )^{2+2\kappa} \right)
\right) \exp\left( \rho \int_{v}^u \frac{ c_{A}}{ \nu^2(\eta) } (
\eta-u)^2 d\eta \right) & v<u.\\
\end{array}\right.
\end{equation*}
Again, the  constants $c^{1,2}$ are such that $\rho=\int_{0}^{1} f(w) dw,
\rho u = \int_{0}^1 w f(w) dw.$ The additional factor in the steady state is due to the fact that the
diffusion is now velocity dependent. If the following relations hold true, both steady states coincide:
\begin{eqnarray}
\sigma(\rho,u) = \rho \lambda, \; \lambda = \lambda_{A} = \lambda_{B}, \; \nu(v)=1, \; 1+2\kappa = \gamma.
\end{eqnarray}
\par

\section{Conclusions}
Starting from microscopic vehicles interactions we derived
Fokker-Planck models for traffic flows as a suitable asymptotic
limit of a Boltzmann dynamic. Among others the new Fokker-Planck
models include the model recently introduced by Illner $\&$ al.
[\ref{Illnerfp}]. A remarkable feature of these models is to
preserve synchronized traffic states and to avoid the degeneracy
in the diffusion term when the vehicle velocity coincides with the
mean velocity. These kinetic models are very promising for
numerical simulation purposes since their computational cost is
strongly reduced when compared to full Boltzmann models. Numerical
results for the non homogeneous multilane case are under progress
and will be presented elsewhere.

%%%%%%%%%%%%%%%%%%%%%%%%%%%%%%%%%%%%%%%%%%%%%
\section*{Acknowledgements}
This work has been supported by DAAD D/06/19582,  DFG HE5386/6-1,
HE5386/8-1, RWTH Seed Funds and Vigoni Project ``Numerical methods
for the simulation and the optimization of traffic flows on
networks''. The first author would like to thank University of
Ferrara for their hospitality.

\medskip
% The data information below will be filled by AIMS editorial staff
Received xxxx 20xx; revised xxxx 20xx.
\medskip

\end{document}